# Hyperspectral terahertz microscopy via nonlinear ghost imaging

**LUANA OLIVIERI[1], JUAN S. TOTERO GONGORA[1], LUKE PETERS[1], VITTORIO CECCONI[1], ANTONIO CUTRONA[1,2], JACOB TUNESI[1], ROBYN TUCKER[1], ALESSIA PASQUAZI[1] AND MARCO PECCIANTI[1,*]**

[1]*Emergent Photonics (EPic) Laboratory, Department of Physics and Astronomy, University of Sussex, Brighton, United Kingdom.*
[2]*Department of Engineering, University of Palermo, Palermo, Italy.*
\* Corresponding author: m.peccianti@sussex.ac.uk

## ABSTRACT
We experimentally demonstrate Time-Resolved Nonlinear Ghost Imaging and its ability to perform hyperspectral imaging in difficult-to-access wavelength regions, such as the Terahertz domain. We operate by combining nonlinear quadratic sparse generation and nonlinear detection in the Fourier plane. We demonstrate that traditional time-slice approaches are prone to essential limitations in near-field imaging due to space-time coupling, which is overcome by our technique. As a proof-of-concept of our implementation, we show that we can provide experimental access to hyperspectral images completely unrecoverable through standard fixed-time methods.

## INTRODUCTION

The reconstruction of complex field distributions in space and time is a challenge in many domains with a significant transversal impact in fields beyond optics, such as microwaves beam steering, ultrasound imaging, and biology [1–8]. On another front, hyperspectral imaging has a pivotal assessment role in many disciplines, as it allows to determine the 2D morphology of an absorption spectrum [9–11]. Hyperspectral imaging assumes a broader probing significance in time-resolved systems; in particular, the delay of each frequency component can be profitably used to access the 3D morphology of the spectral phase response of a target, i.e., its spatially resolved complex dielectric function. Modern photonic approaches have produced essential breakthroughs in medicine, biology, and material science imaging [11–15]. In this context, the ability to reconstruct the time-domain waveforms provides direct access to the field [16], although these approaches are well established in microwave and ultrasound imaging [2,5,6], they are sensibly less diffused in photonics. Terahertz (THz), in this regard, has emerged as one of the most relevant photonics frameworks in which the time-evolution of a field amplitude is experimentally accessible. Indeed, THz Time-Domain Spectroscopy (TDS) has played a pivotal role in establishing THz as an independent research field [17–20].

Single-pixel imaging approaches find their origin in domains where single-point detectors outperform detector-arrays in terms of specifications or availability [21,22], for this reason, they have attracted interest also in the THz community [23–25]. In photonics, these methods have unlocked the powerful ability to add multiple dimensions and novel functionalities to simple spatial probing, enabling several breakthroughs in classical and quantum imaging [26–32]. In its most modern connotation, the Ghost-Imaging (GI) is a form of computational imaging that employs the sequential illumination of an object with a set of pre-determined patterns [22,33–37].

In terms of accessing new emerging wave domains, such as THz, the GI offer the option of closing relevant technological gaps while raising new challenges, such as the limited availability of THz Spatial Light Modulators (SLMs) and the coarse diffraction limit [26].

The combination of single-pixel imaging and TDS provides the exciting possibility to exploit novel space-time computational imaging approaches [31,32], and the TDS-GI has been recently proposed as viable for THz imaging [38–42]. The THz field can be densely sampled in space, giving access to sub-wavelength microscopy when an object is exposed to the near field of a THz source, detector, or mask. Besides its potential practical impact in THz microscopy, GI microscopy provides an accessible fundamental framework for investigating time-resolved imaging in the presence of strong spatiotemporal coupling, a dominant condition in the near-field domain.

In this article, we experimentally implement an imaging protocol based on the Time-Resolved Nonlinear Ghost Imaging (TNGI), which we have recently theoretically proposed as a single-pixel imaging method where a set of nonlinear wavelength transformations are inserted in both the illumination and detection chains [43]. We generate the THz patterns used for the GI reconstruction by nonlinear conversion of spatially modulated optical pulses in a quadratic medium. Leveraging on the time-dependent field detection, as opposed to the intensity detection usually implemented in optics GI-equivalent, we implement the detection in the Fourier plane, effectively acquiring the average value of the scattered field. With this approach, the system resolution is effectively independent of the numerical aperture of the detection system, in sharp contrast with standard single-pixel approaches working in optics. We test our time-dependent THz microscope on benchmark images, showing the capability of our system to extract the spectral morphologies, such as the water content in a leaf.

Most importantly, we demonstrate near-field, coherent hyperspectral imaging in a regime where spatiotemporal coupling is strongly evident. We experimentally show that, in this regime, the image information is inherently inaccessible when the reconstruction is performed at fixed time-slices of the transmitted field, as the traditional iso-time imaging approaches become affected by errors and artifacts. We show experimentally that in near-field the full spatiotemporal signal is required to preserve space-time imaging, and we provide a methodology, which we refer to as 'space-time refocusing' for high fidelity reconstruction. Interestingly, we also show experimentally that the thickness of the generation crystal does not preclude significantly higher resolutions (as in some of the proposed THz-GI approaches).

**METHODS: THE TIME RESOLVED NONLINEAR GHOST-IMAGING**

We formulated the TNGI as a single-pixel imaging approach based on the time-resolved detection of the electromagnetic field scattered by a sample, as opposed to the standard formulation of GI that relies on the time-averaged field intensity [43]. Without loss of generality, the TNGI describes the optical and morphological features of a sample through a spatiotemporal transfer function $T_{sample}(x, t)$ which is reconstructed through a sequence of measurements as follows:

$$T_{sample}(x, t) = \langle C_n(t) I_{opt}(x) \rangle_n - \langle C_n(t) \rangle_n \langle I_{opt}(x) \rangle_n, \qquad (1)$$

where $I_{opt}(x,y)$ is the intensity distribution of the $n$-th incident optical pattern, and $\langle\cdots\rangle_n$ is the average over the distribution of patterns. In Eq. (1), the expansion coefficients $C_n(t)$ are defined as

$$C_n(t) = \int d\mathbf{x}\, E_n^+(\mathbf{x},t), \qquad (2)$$

and correspond to the spatial average of the electric field $E_n^+(\mathbf{x},t)$ transmitted by the sample and acquired by TDS detection (see Supplementary Section S3). Note that Eq. (1) is closely related to the linear formulation of the standard GI where the incident and scattered intensities are linearly related. Such a similarity is a direct consequence of the optical-to-THz conversion taking place in quadratic media, where the generated THz field is expressed as

$$E_{THz}(x,y) \propto \chi^{(2)} I_{opt}(x,y) \qquad (3)$$

where $\chi^2$ is the second-order nonlinear susceptibility of the nonlinear medium[1]. The capability of directly controlling the THz field by acting on the incident optical intensity is an essential feature of our approach, as in casting Eq. (1) we do not require any assumption stemming from the binary nature of the illumination (as required, e.g., in mask-based GI [39–42]). It is also worth noticing that, differently from the standard GI formulation, the coefficients in Eqs. (1-2) are built up by coherent measurements of the electric field, and they do not represent the scattered intensity.

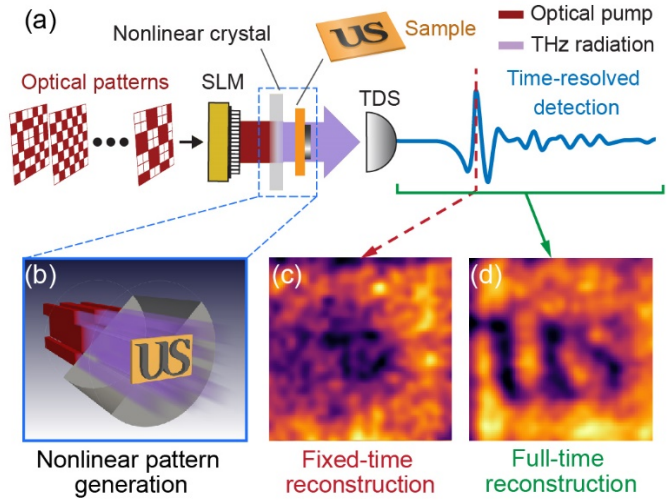

**Figure 1: Conceptual description of the Time-resolved Nonlinear Ghost Imaging (TNGI) approach.** (a) Key experimental components and methodology. (b) Volumetric representation of the nonlinear generation of THz patterns. (c) Fixed time reconstruction with field of view 2mm×2mm and 32×32 spatial sampling. (d) Backpropagated hyperspectral image, averaged between 1 and 2 THz.

The principal elements of our experimental implementation of the TNGI are shown in Fig. 1. We impressed a series of intensity patterns on an ultrafast optical beam (λ=800 nm, repetition rate 1 kHz, pulse duration 75 fs) using a commercial wavefront-shaping device. In our experiments, we employed both a binary amplitude digital-micromirror-device (DMD) and a phase-only Liquid Crystal on Silicon (LCoS) spatial light modulator. We converted the optical pattern to a THz field distribution $E_{THz}(x,y)$ through nonlinear optical rectification in a quadratic crystal (ZnTe) of thickness $z_0$. The generated THz pattern sampled different targets (in our experiments different metallic masks and dielectrics) placed in proximity of the crystal surface, and the average transmitted field was measured through Electro-Optic (EO) detection. In our THz implementation, the $C_n(t)$ coincide with the electric field detected via Time-Domain-Spectroscopy (TDS) at the center of the Fourier plane (i.e., at $k_x=k_y=0$) [44]. As an image-reconstruction protocol, we exploited a Walsh-Hadamard encoding scheme (with "Russian Doll" ordering [45]) based on binary amplitude patterns, which is known to maximize the SNR of single-pixel imaging schemes [39]. A detailed schematic and further details on the optical setups are included in the Supplementary Information (SI).

---
[1] $E_{opt}^*(x,y)$ stands for the complex conjugate field.

The use of nonlinear conversion to generate THz patterns provides a series of features when developing a single-pixel TDS imaging scheme. First, the ability to control the THz field distribution by shaping the optical field, as expressed by Eq. (3), allows us to generate patterns with subwavelength resolution when compared to the THz wavelength (300μm, at 1THz). The resolution of the optical pattern $I_{opt}(x, y)$ is ultimately bound by the optical diffraction limit and the numerical aperture of the optical setup. Secondly, the SNR of the detected THz signal increases linearly with the incident optical intensity. Finally, there is perfect temporal coherence and spatial locality between the pump pulse and the distribution of THz sources. Temporal and spatial coherence is a direct consequence of the nonlinear conversion process and has significant consequences on our capability of imaging samples in challenging experimental conditions. An open issue in THz-GI concerns the impact of the distance between the THz pattern source and the sample, as required when assessing the transmission from a sample placed on a holding substrate. As discussed in [39,43], the near-field propagation of subwavelength patterns exhibits spatiotemporal coupling between adjacent pixels, altering the spatial and temporal features of the pattern. Under these conditions, the sampling function impinging on the object is not the original, pre-determined pattern, but its space-time "propagated" version. Such discrepancy introduces a systematic and ineliminable error in determining the scattered waveform from the object using a time-sliced (or iso-time) imaging. While the effect of spatiotemporal coupling could be reduced when the sample distance from the sources is much smaller than the resolution targeted (i.e. by employing a thin patterning substrate), the error introduced by diffraction (and by the interaction with samples with complex transmission properties) is always present. Such an error is not quantifiable in the case of single time-slice acquisition, and it cannot be represented by standard definitions of signal-to-noise ratio employed in image analysis. The combination of optical coherence and direct field detection allows us to reverse the effects of spatiotemporal coupling, to obtain the correct time-domain reconstruction of a sample within one wavelength of distance, and to perform coherent hyperspectral imaging through TNGI.

**EXPERIMENTAL RESULTS: HYPERSPECTRAL IMAGING**

As a first case study, in Figure 2 we present the 2mm×2mm spatiotemporal image of a metallic structure deposited on a 50 μm Kapton substrate. The image was retrieved by shaping the optical illumination with a binary DMD and by placing the metallic structure in the proximity of a $z_0=1$ mm thick ZnTe generation crystal. We achieved analogous results using an LCoS modulator, as shown in Supplementary Figure 2. By retrieving the spatiotemporal image of the sample (Fig. 2a), we can capture the full effects of the interaction between the THz field and its subwavelength metallic features. As can be observed at $t = 0.2$ ps, in fact, the object does not appear just as a blocking mask, but it features complex resonances from the edges of the object. The time-resolved measurement of the scattered field allows us to reconstruct also the hyperspectral image of the sample (Fig. 2b). Interestingly, in our experiments, we were able to resolve features within the 50-100µm scale even in the presence of a relatively thick generation crystal (as opposed to the typical thickness requirements in other approaches [39–42]). The field-spatial spectra of each generating layer in the ZnTe do not mix incoherently (see Supplementary Section S2).

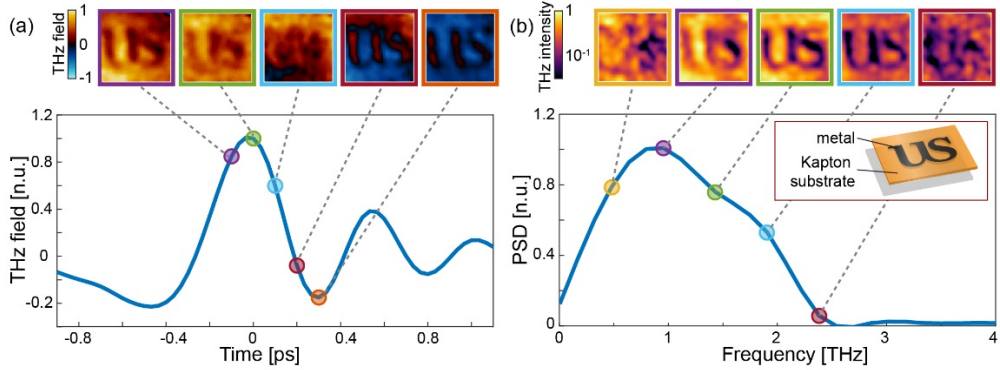

**Figure 2: Spatiotemporal image of a metallic sample.** (a) Temporal evolution of the metallic sample with fixed times image reconstructions. (b) Spectral response with hyperspectral images. The field of view was 2mm×2mm with a 16×16 spatial sampling.

Coherent access to the temporal field evolution allowed us to reconstruct full spatiotemporal images of semi-transparent samples. As a relevant example (and to credit a similar image in [19], widely considered one of the first milestones in THz imaging), we show in Figure 3 an image of a leaf at different stages of desiccation. As can be evinced from Fig (3c-d), the iso-time image of a semi-transparent sample is significantly harder to interpret than a standard metallic mask, as the different parts of the sample induce different temporal delays and phase shifts. Nevertheless, we were able to retrieve the TDS time-trace in different points of the sample (Fig. 3e,h) and retrieve its hyperspectral image both in terms of amplitude (Fig. 3f,i) and spectral phase (Fig. 3g,j), allowing us to reconstruct its morphology and spectral fingerprint. As relevant examples, we present data for a fresh leaf (Fig. 3e-g) and a dried leaf (Fig. 3h-j). By comparing the transmission from the two samples, it is possible to assess the changes in water content as in [19].

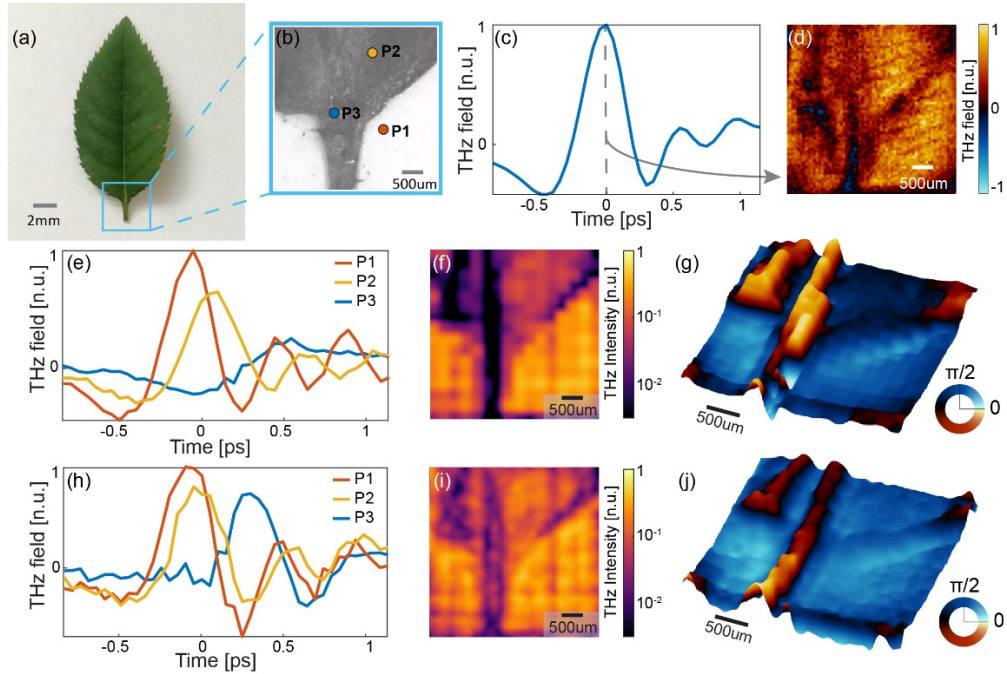

**Figure 3: Hyperspectral image of a leaf.** (a) Optical image of the leaf. (b) Microscope image. (c) Temporal evolution of the leaf. (d) Fixed time reconstruction (128×128 pixels). (e) Local temporal evolution of the fresh leaf in the points indicated in (b). (f) Hyperspectral image of a fresh leaf at 1.5THz (16×16 pixels). (g) Phase image of the fresh leaf. (h-i-j) Same as the previous panel for a dried leaf. (32×32 pixels images). All the images correspond to a field of view of 4mm×4mm.

## IMAGING THROUGH INVERSE-PROPAGATION

The experimental results presented in Figure 4 explore a relevant consequence of the space-time coupling in near-field TNGI. In this case, we collected the image of a metallic sample, analogous to the one in Fig. 2, but introducing a non-negligible distance between the sample and the emitter, which includes the Kapton substrate, in a typical time-of-flight imaging case.

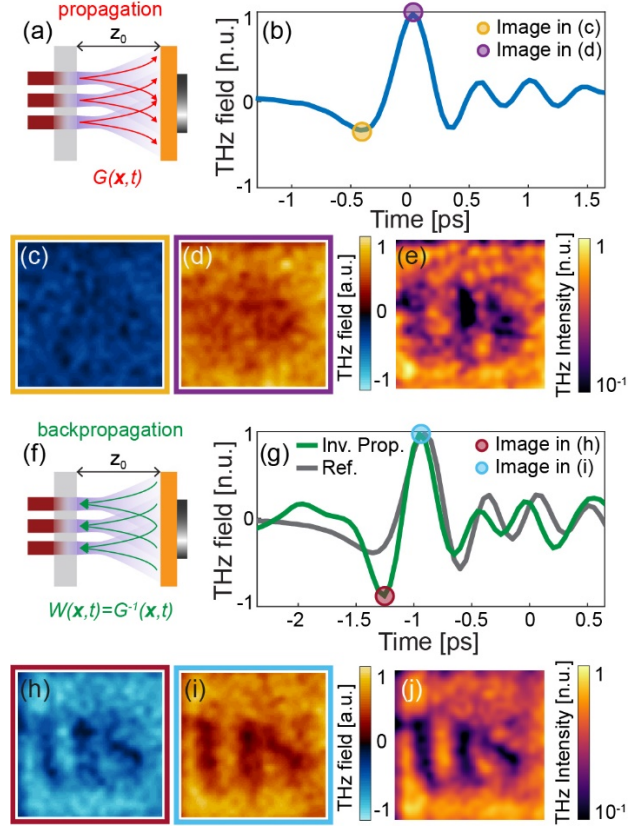

**Figure 4: Time-resolved image reconstruction: inverse-propagation approach.** (a) Conceptual illustration of the propagating imaging scheme: the sample is placed at $z_0=300\mu m$ from the crystal. (b) Temporal evolution of the sample. (c-d) Fixed time reconstructed images at the points indicated in (b). (e) Hyperspectral image averaged between 1 and 2 THz. (f) Conceptual illustration of the backpropagation scheme. (g) Temporal evolution of the backpropagated image (green) and the temporal evolution without the sample (grey). (h-i) Fixed time reconstruction of the backpropagated image at the points indicated in (g). (j) Backpropagated hyperspectral image, averaged between 1 and 2 THz. In all panels the field of view was 2mm×2mm with a 32×32 spatial sampling.

In these conditions, the sample morphology cannot be appreciated in any of the iso-time images regardless their temporal position (Fig. 4c, d show some examples), or in the hyperspectral image (Fig. 4e), which shows quite distorted image even if in some pixels a high contrast is reached. As theoretically demonstrated in [43], such a limitation is a direct consequence of spatiotemporal coupling, which leads to a substantial modification of the incident sampling patterns as they propagate (Fig. 4a). At this stage, it should be observed that our TNGI protocol relies on the collection of the average field as performed by sampling the origin of an optical Fourier plane (i.e., $k_x = k_y = 0$). As a result, it only requires an optical system capable of collecting a very narrow spatial spectrum and the numerical aperture of the imaging system plays a minimal role in defining the image resolution. On the contrary, the SNR of the THz detection plays a fundamental role in resolving the vanishing near-field scattered contributions (at high spatial frequencies) for increasing values of the distance between the sample and the emitter.

With the sensitivity available, we could then backpropagate the pattern sampling function in order to "space-time refocus" the image (Fig. 4f) and reverse the effect of spatiotemporal coupling (see Supplementary Section S3 for a theoretical discussion on the inverse-propagation reconstruction) [43]. This procedure allows us to retrieve the correct time-resolved image of the scattered field in the proximity of the sample, restoring the morphological and spectral features of its hyperspectral image (Fig. 4i-j). We argue that the inverse-propagation reconstruction is a strict requirement to reconstruct the sample properties at different depths, i.e. in near-field time-of-flight imaging.

## 5. Conclusions

In conclusion, we performed the first experimental example of the Time-resolved Nonlinear Ghost Imaging approach exploiting a nonlinear quadratic conversion. We devised a general reconstruction method based on the linear dependence between impinging optical patterns and the detected THz time-domain field average. The approach enables hyperspectral imaging as performed in the state-of-the-art by TDS imagers. It features near-field imaging and shows relaxed constrains in terms of thickness of the nonlinear converter (our proof-of-concepts exploits off-the-shelves nonlinear substrates). As predicted in [43], we demonstrated that popular iso-time approaches are not suitable for near-field spatiotemporal microscopy and this a central issue when an object comprises elements at different optical depths. We proved experimentally that thanks to the spatial and temporal coherence, it is possible to devise an inverse-propagation operator capable of "refocusing" the image in the space-time and, therefore, correctly reconstructing the hyperspectral image of the sample. We believe this work can have a substantial impact in the field of near-field imaging, especially in light of the emergence of thinner and more efficient THz emitters (e.g., spintronic substrates, surface emitters or novel materials with exceptionally high nonlinear coefficients such as DSTMS crystals) [46–49].

**Funding**
This project has received funding from the European Research Council (ERC) under the European Union's Horizon 2020 research and innovation programme (Grant agreement n° 725046), from the UK Engineering and Physical Sciences Research Council EP/N509784/1 and Industrial Innovation Fellowship Programme, under Grant EP/S001018/1.

**Acknowledgments**
J.T. acknowledges the support of the EPSRC through the studentship EP/N509784/1. J.S.T.G, L.P., V.C., and R.T. acknowledge the support of the European Union's "Horizon 2020" research and innovation program, grant agreement no. 725046, ERC-CoG project TIMING. J.S.T.G. acknowledges funding from the Helena Normanton Fellowship of the University of Sussex, UK. A.C. acknowledges the support from the University of Palermo, Italy through the mobility fellowship "Corso di Perfezionamento all'estero 2018".

**Disclosures**
The authors declare no conflicts of interest.

**SUPPORTING MATERIALS**
See Supplementary Materials for supporting content.